\documentclass[a4paper]{scrartcl}

\usepackage{amsmath} 
\usepackage{graphicx} 
\usepackage{authblk}
\usepackage{xcolor}
\usepackage{siunitx}
\usepackage{subcaption}
\usepackage{chemformula} 

\title{Study of NbN as superconducting material for the usage in superconducting radio frequency cavities}

\author[1]{Kristof Schmieden}
\author[1]{Tim Schneemann}
\author[2]{Matthias Schott}
\author[1,3]{Malavika Unni}
\author[1,3]{Hendrik Bekker}
\author[1,3]{Arne Wickenbrock}
\author[1,3]{Dmitry Budker}
\affil[1]{Johannes Gutenberg Universit\"at Mainz, 55128 Mainz, Germany}
\affil[2]{Rheinische Friedrich-Wilhelms-Universität Bonn, 53113 Bonn, Germany}
\affil[3]{Helmholtz-Institut Mainz, GSI Helmholtzzentrum f\"ur Schwerionenforschung GmbH, 55128 Mainz, Germany}

\begin{document}

\newcommand{\QMax}{\ensuremath{3.46\cdot 10^5}}
\newcommand{\todo}[1]{\textcolor{red}{\bfseries #1}}
\newcommand{\abs}[1]{\lvert #1 \rvert}
\newcommand{\Tc}{\ensuremath{10.34\pm 0.24}~\text{K} }
\newcommand{\fres}{8.4~GHz }

\maketitle

\begin{abstract}

A new axion-haloscope is setup at the Johannes Gutenberg university of Mainz, named the \textsc{Supax} (a SUPerconducting AXion search) experiment. 
This setup is used to characterize the behaviour of a NbN coated superconducting cavity in a 2.5~T strong magnetic field, at a resonance frequency of \fres. 
We observe an increasing surface resistance with increasing magnetic field, leading to a decreasing quality factor. The behaviour is similar to that of previously studied cavities using \ch{Nb3Tn}. 

\end{abstract}

\section{Introduction}
The quest to find the axion solving the strong-CP problem \cite{PhysRevLett.38.1440,PhysRevLett.40.223,PhysRevLett.40.279} has spawned a multitude of experiments, which exploit coupling of the axion field to nucleons, gluons and photons. A detailed overview on the experimental searches for axions can be found in Refs. \cite{Graham:2015ouw, Adams:2022pbo}. 
Three types of experiments study the axion-photon coupling and are targeting different sources of axions: 
Light-shining-through-wall experiments (LSTW), where the axions are produced in the laboratory; helioscopes using the sun as source and haloscopes looking for axions 
in the dark matter halo of our galaxy \cite{Sikivie:1983ip}. 
In all three types of experiments the axions convert in a 
magnetized volume via the inverse Primakoff effect into photons. The photon frequency corresponds to the axion energy.
To boost the conversion rate of axions to photons both LSTW and haloscope experiments utilize resonators. For typical haloscope experiments the 
resonance frequencies lie in the radio frequency (RF) regime and cavities 
are used to resonantly enhance the signal.
The signal over noise ratio of a haloscope experiment is given by
\begin{equation}
    \text{SNR} \propto g_{a\gamma\gamma}^2 m_a V C_{mnl} B^2 Q_L \frac{1}{T_{\text{sys}}}
\end{equation}
where $g_{a\gamma\gamma}$ is the axion--photon coupling, $m_a$ the axion mass, $B$ the static magnetic field, $V$ the conversion volume of the cavity, $T_{\text{sys}}$
the systems noise temperature and $C_{mnl}$ the geometric form factor describing the overlap of the cavity mode with the magnetic field. $Q_L$ is the loaded quality factor of the cavity and the main topic of this paper. 

In recent years superconducting cavities were studied by several experiments like RADES \cite{Golm:2021ooj} and at CAPP \cite{Ahn:2021fgb} in order to increase the quality factor and hence the sensitivity. 
As it turns  out, classic superconductors, in particular pure \ch{Nb} and \ch{Nb3Tn} exhibit an increasing surface resistance with increasing magnetic field and show lower quality factors than non superconducting copper cavities at high magnetic fields. 
High temperature superconductors (HTS) based on rare-earth barium copper oxide (REBCO) show good performance in magnetic fields with quality factors above $3\cdot 10^5$ at 8\,T field as shown in \cite{Ahn:2021fgb}.
However, HTS tapes are not suitable for coating curved surfaces significantly constraining usable cavity geometries. 
REBCO coated cavities are typically built out of many pieces with flat surfaces, approximating the desired geometry. This works well for cylindrical shapes, where the simulated losses due to not coated ends of the cavity are in the order of $10\%$ \cite{Golm:2021ooj}.
For other shapes, like spherical cavities a significant reduction in performance is expected in addition the mechanical challenges. 
To overcome this problem we study niobium nitride (NbN) as superconductor in the context of a new axion haloscope experiment, \textsc{Supax}, which is currently prepared \cite{Schmieden:2021msb} at Mainz and Bonn. 

\section{Cavity Design}
\begin{figure} [ht]
    \centering
    \includegraphics[width=0.245\textwidth, angle=90]{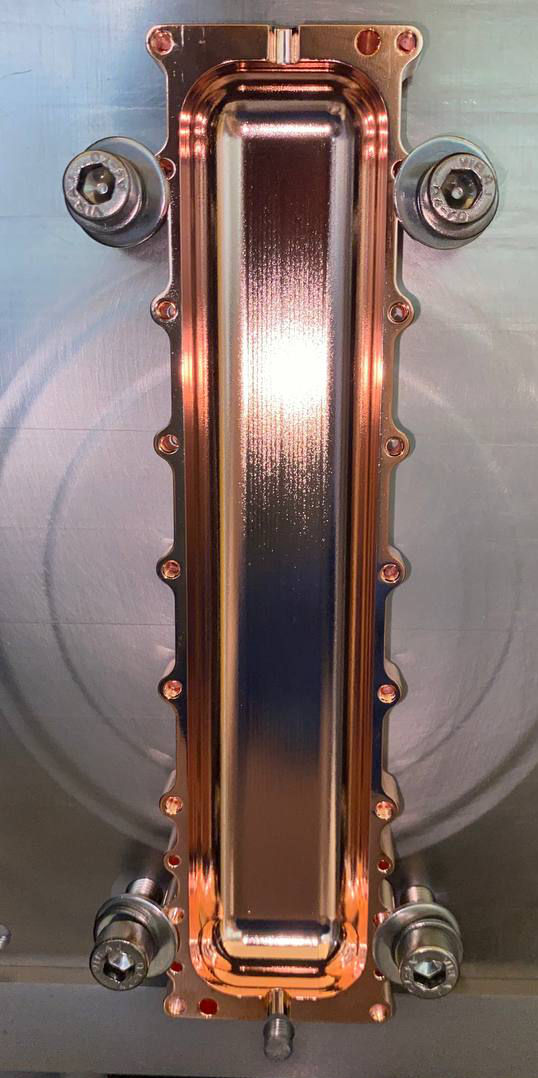}
    \hfill
    \includegraphics[width=0.245\textwidth, angle=90]{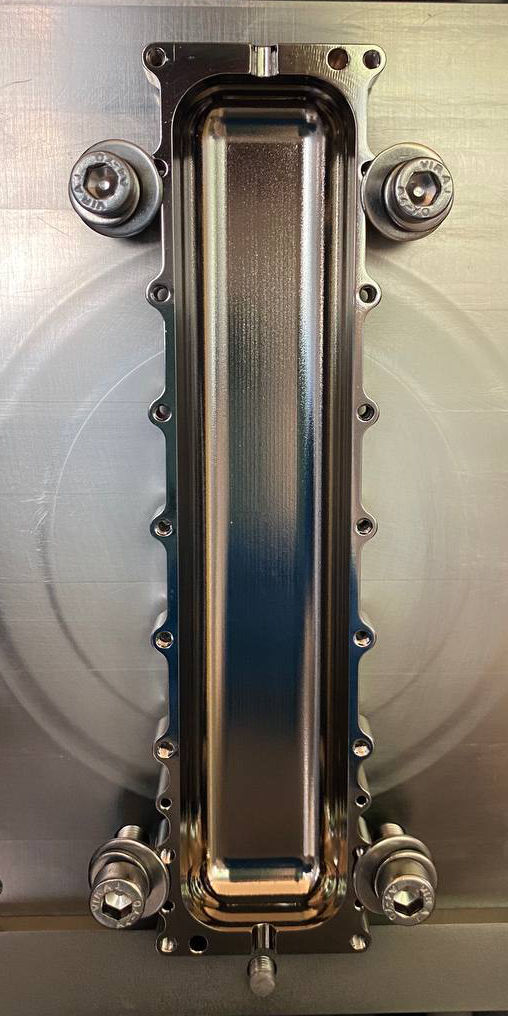}
    \caption{Images of the cavity under test. Left: the electropolished copper cavity. Right: after depositing the NbN superconductor.}
    \label{fig:cavity}
\end{figure}

A cavity with a resonance frequency of the TM$_{010}$ mode of 8.4~GHz was designed, based on experiences from the RADES group \cite{Golm:2021ooj}. 
The cavity is built out of two copper half-shells with dimensions 27~mm x 40~mm x 160~mm. The inner dimensions are 22.8~mm x 30~mm x 150~mm where all corners of the cavity are rounded with a radius of 9~mm. 
Expendable material is removed from the outside as much as possible to lower the heat capacity. 
A $2\,\mu$m thick layer of NbN was deposited on the copper substrate \cite{Leith_2021, ubsi_2108}. 
Images of the cavity half-shell before and after the coating are shown in Fig. 
\ref{fig:cavity}. Two antennas are attached to the cavity at opposite ends of its long-axis. One is strongly coupled to the TM$_{010}$ mode, the other weakly. 
The resonance frequency of the TM$_{010}$ mode is measured to be \fres at a temperature of 4~K.

\section{The Measurement Setup}
\begin{figure}[t]
    \centering
    \begin{minipage}[t]{0.49\textwidth}
    \includegraphics[width=0.99\textwidth]
    {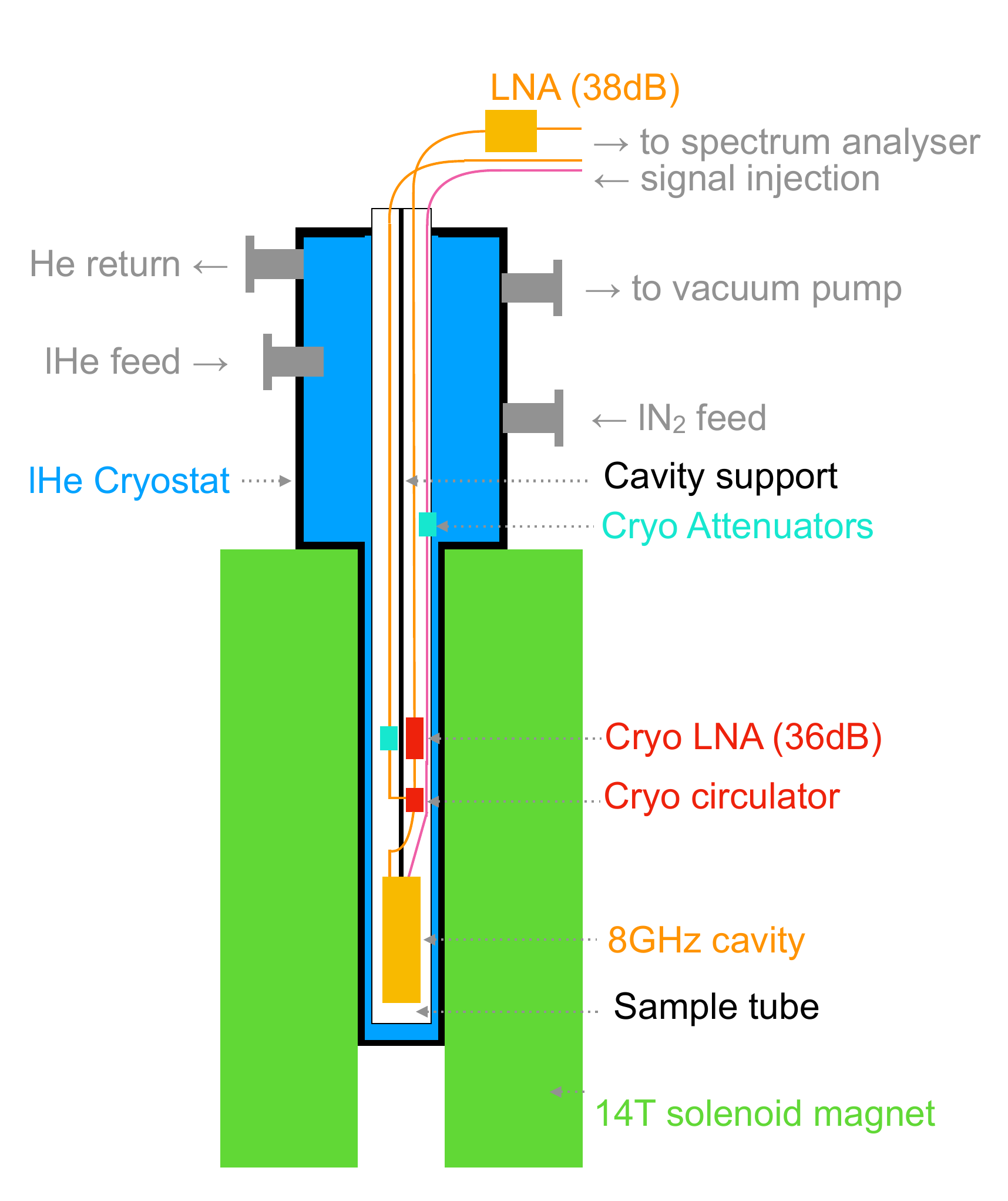}
        \caption{Schematic drawing of the setup of the \textsc{Supax} experiment.}
        \label{fig:setup}
    \end{minipage}\hfill
    \begin{minipage}[t]{0.49\textwidth}
    \centering
    \includegraphics[width=0.71\textwidth]{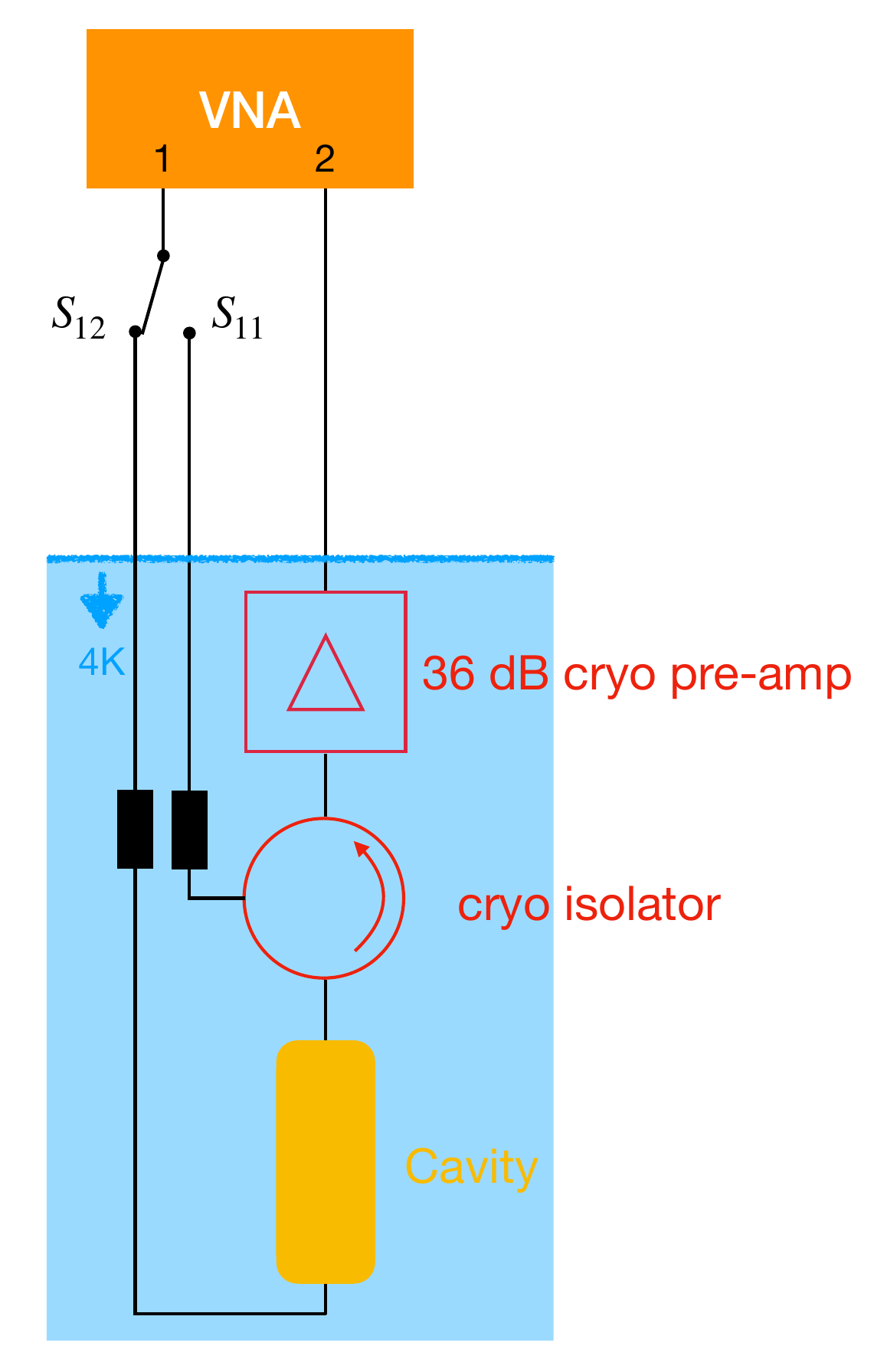}
        \caption{Schematic view of the data acquisition system.}
    \label{fig:setupDAQ}
    \end{minipage}  
    
\end{figure}

The \textsc{Supax} experiment is currently setup at the University of Mainz and utilises a liquid helium cryostat inserted into the room temperature bore of a 14.1~T magnet located at the Helmholz-Institute Mainz. 
A schematic drawing of the experiment is shown in Fig. \ref{fig:setup} together with its data acquisition system in Fig. \ref{fig:setupDAQ}.

The temperature of the cryostat can be adjusted between 1.5~K and 300~K as needed and is stabilised to better than 0.1~K at 4~K. The magnetic field can be adjusted between 0 and 14.1~T with a ramp time of approximately 5\,h. 

Signals from the cavity are amplified by a 36~dB cryogenic low noise amplifier, which is connected via a circulator to the strongly coupled cavity port to avoid reflections off the amplifier. 
Signals can be injected into the cavity either via the weakly coupled antenna or through the circulators 3rd port and the strongly coupled antenna. 
Both input lines are equipped with attenuators which are thermally coupled to the base temperature of the cryostat to diminish thermal noise from outside of the cryostat.
This setup allows measuring the $S_{11}$ and $S_{12}$ parameters with a vector network analyser without changing the setup. 
 
\begin{figure}
\centering
\hfill
\begin{subfigure}[b]{0.48\textwidth}
\includegraphics[width=\textwidth]{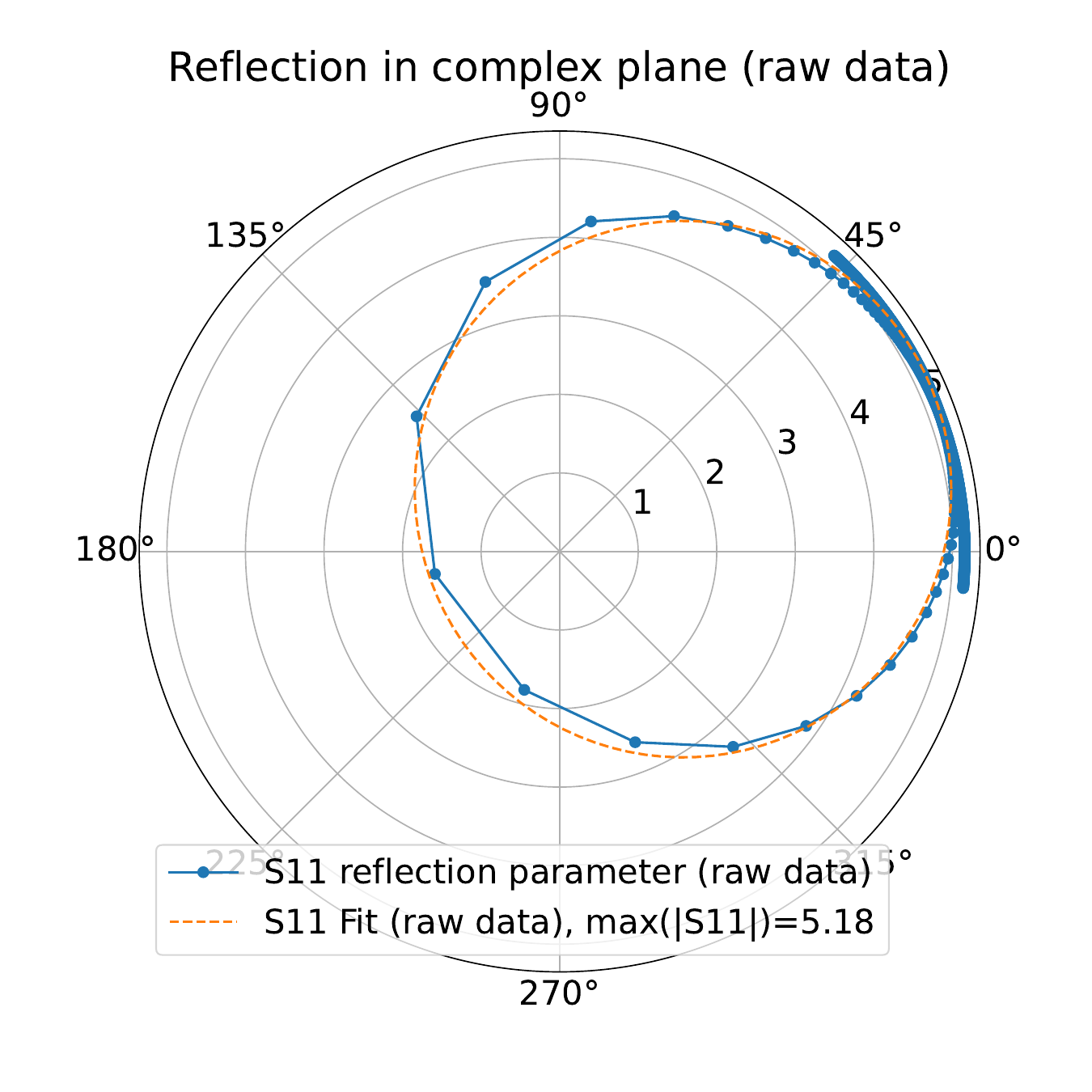}
\label{fig:circle}
\end{subfigure}
\hfill
\begin{subfigure}[b]{0.48\textwidth}
\includegraphics[width=\textwidth]{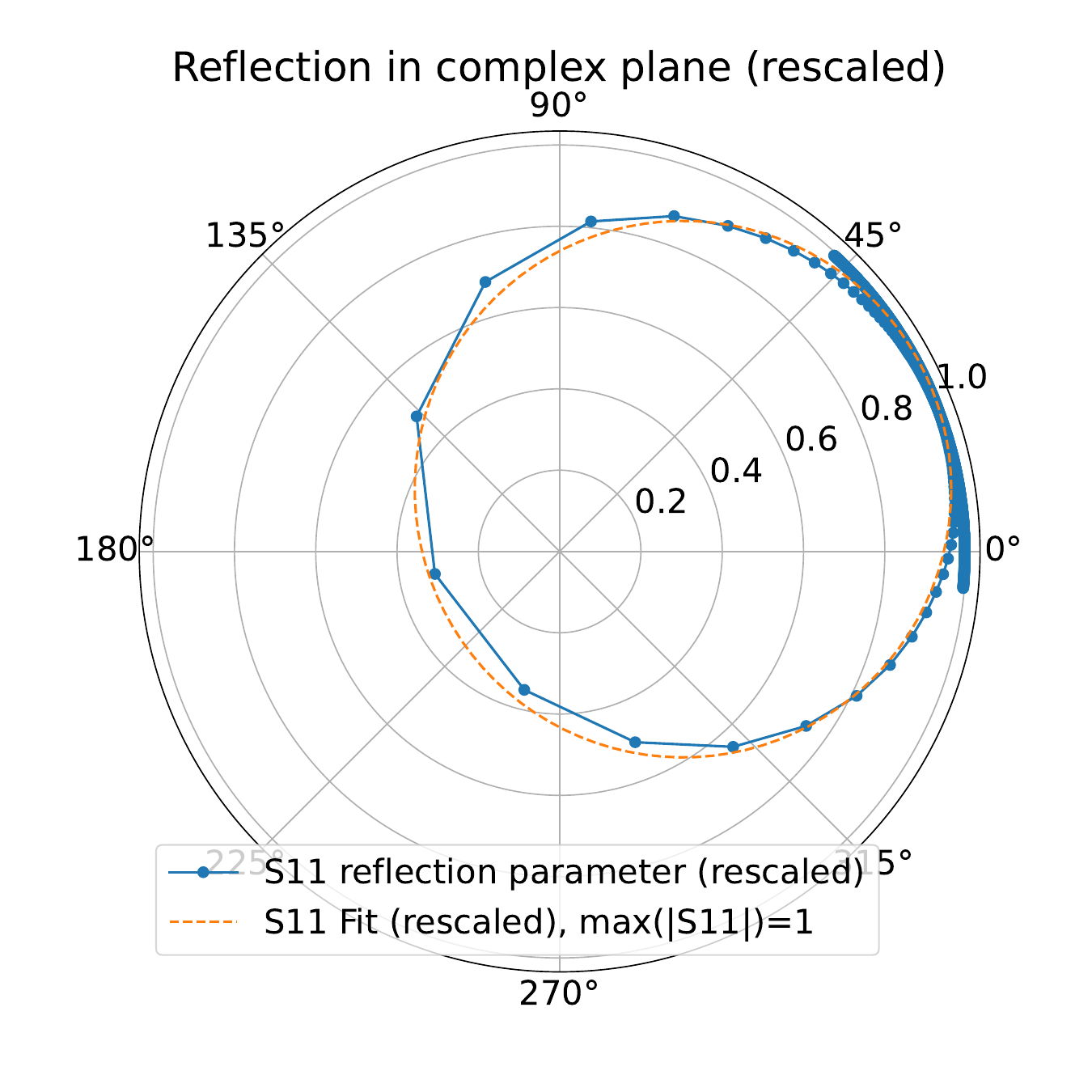}
\label{fig:circle_rescaled}
\end{subfigure}
\caption{On the left is the raw $S_{11}$ data plotted on a complex plane. On the right is the rescaled $S_{11}$ data so that the maximum of the circle fit has $\abs{S_{11}} = 1$ ensuring that in the peak region amplification (from pre-amplifier) and cable losses are accounted for.}
\label{fig:circle_fit}
\end{figure}

The resonance frequency of the cavity is determined from the $S_{21}$ parameter to be $f_0 = 8.4$~GHz.
The loaded quality factor $Q_L$ is calculated dividing the centre frequency by the width of the resonance curve $3\,\text{dB}$ below the maximum.
\begin{equation}\label{eq:QL}
    Q_L = \frac{f_0}{\text{FWHM}}.
\end{equation}
Due to the installed pre-amplifier and losses in the cables the normalization of the measured S-parameters is lost and needs to be restored. 
For this the $S_{11}$ parameter is plotted in the complex plane, where a scan over the resonance peak will form a circle. 
The data is rescaled so that the fit of the data around the peak has its maximum magnitude at $\abs{S_{11}} = 1$, as shown in Fig. \ref{fig:circle_fit}.
The rescaled data is converted to a linear scale and the coupling factor $\beta$ is determined as
\begin{equation}\label{eq:beta_undercoupled}
    \beta = \frac{1-\min(\abs{S_{11}})}{1+\min(\abs{S_{11}})}
\end{equation}
if the port is undercoupled and
\begin{equation}\label{eq:beta_overcoupled}
    \beta = \frac{1+\min(\abs{S_{11}})}{1-\min(\abs{S_{11}})}
\end{equation}
otherwise. Note that for critical coupling ($\beta = 1$) eqs. \ref{eq:beta_undercoupled},\ref{eq:beta_overcoupled} are equal since $S_{11}^{\beta=1} = 0$ at the resonance frequency. The information if the port is over- or under coupled is obtained from the fit of a circle to the rescaled $S_{11}$ data. 
If the diameter of the circle is larger than 1, the port is overcoupled and undercoupled otherwise.

\section{Results}
\begin{figure}[th]
    \centering
    \begin{minipage}[t]{0.49\textwidth}
        \includegraphics[width=0.95\textwidth]{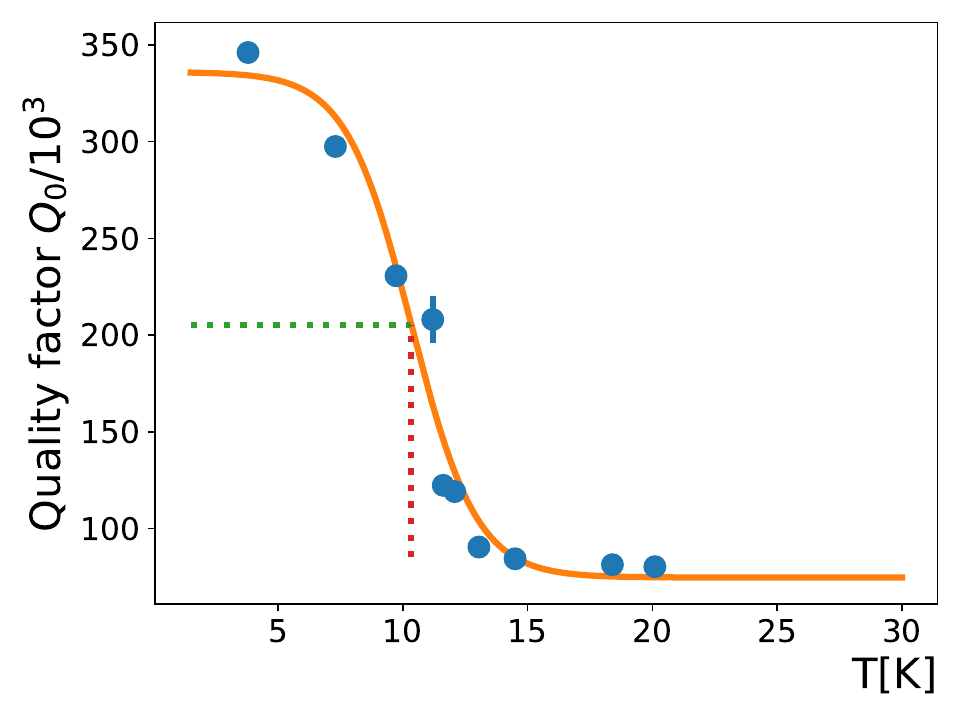}
        \captionof{figure}{Temperature vs. quality factor of the NbN coated cavity. A critical
        temperature of $T_c = \Tc$ is measure with a maximum quality factor of \QMax at 4~K. The dashed lines indicate the critical temperature where the sigmoid fit function reaches half it's max. value.}
        \label{fig:results_Tc}
    \end{minipage}\hfill
    \begin{minipage}[t]{0.49\textwidth}
       \includegraphics[width=0.95\textwidth]{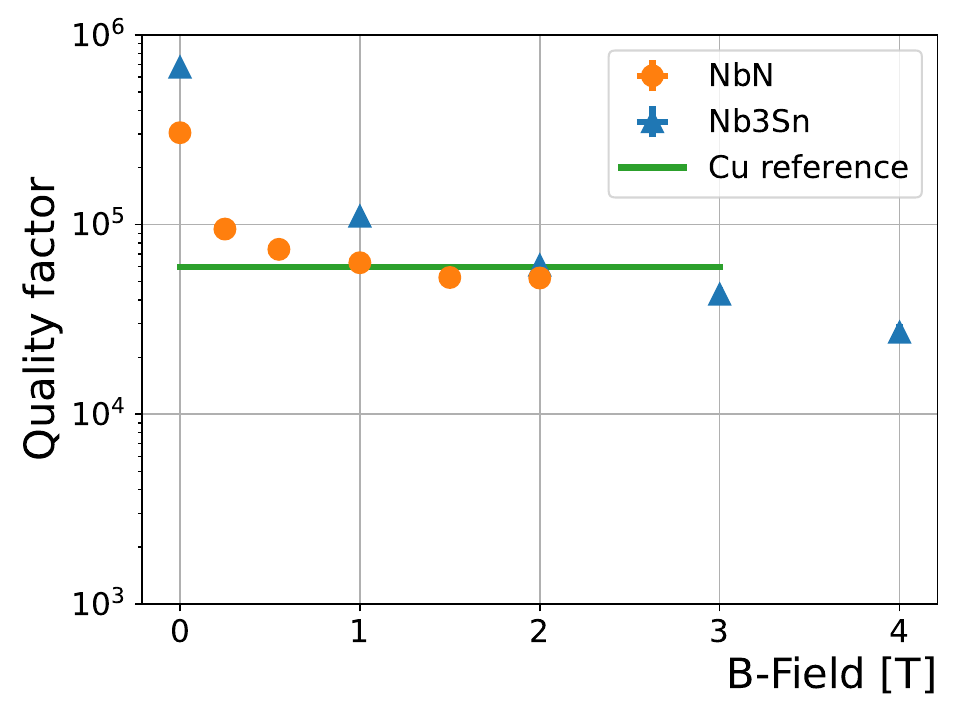}
        \caption{
        Dependence of the maximum quality factor $Q_0$ at 4~K on the external magnetic field strength. For comparison the behaviour of \ch{Nb3Tn}, taken from \cite{Golm:2021ooj}, is shown as triangles and the value for a non-coated Cu cavity as line. 
        }
    \label{fig:results_B-field}
    \end{minipage}  
\end{figure}

In a first measurement the transition curve of the NbN coating is measured without magnetic field and the result is shown in Fig. \ref{fig:results_Tc}.
A sigmoid function plus a constant is fitted to the measured values. The transition temperature $\textrm{T}_c$ taken to be the temperature where the sigmoid reached half it's  maximum value, yielding $\textrm{T}_c = \Tc$. The maximum quality factor at 4~K
is measured to be \QMax, a factor of five above the value of the non-coated copper cavity. 

The dependence of the quality factor on the magnetic field is measured by heating the cavity well above the transition temperature $T_c$ to 20~K, adjusting the magnetic field to the desired value and cooling the cavity again to 4~K. 
The system is then stabilized at 4~K for several minutes before measuring the S-Parameters and subsequently repeating the procedure. The loaded and unloaded quality factors are then determined from the measured S-Parameters, as described above.

Two sources of uncertainties are considered: the uncertainty on the measurement of the loaded quality factor $\Delta Q_L$ and the uncertainty on the determination of the antenna coupling $\Delta \beta$. 
$Q_L$ is determined from the $S_{12}$ parameter as shown in eq. \ref{eq:QL}. As uncertainty on the measured resonance frequency $f_0$ half a bin-width of the VNA measurement is assumed, corresponding to $\pm 5$~kHz. The corresponding uncertainty on the FWHM estimate is $\pm7.1$~kHz. 
The determination of the coupling coefficient $\beta$ relies on the measurement of $\min(|S_{11}|)$ after rescaling, which is estimated by fitting a modified Lorentzian function to the measured data. Changing the fit range by $\pm 20\%$ has a negligible impact on the fit result compared to the uncertainty on the fit parameters.
The uncertainty ranges from $0.72\,\% \le \Delta\beta \le 2.28\,\%$.
The resulting uncertainty on the unloaded quality factor $Q_0$ is indicated as error-bars on the results presented in Fig. \ref{fig:results_B-field} and around 2~\%.
The uncertainty on the temperature of the cavity is $< 0.1$~K and hence neglected. 
The measured $Q_0$ drops with increasing external magnetic field and surpasses the value of the reference copper cavity at $B = 1.5\,\text{T}$. The observed behaviour is similar to that of \ch{Nb_3Sn}. This result shows that a $2\,\mu$m thick NbN coating is not beneficial when operating a cavity in a static magnetic field above $1.5\,$T. Other promising candidates for a coating are iron based superconductors, like \ch{FeSn}.




\section*{Acknowledgement}
\label{sec:acknoledgement}

We thank Jessica Golm for many helpful discussions and for providing the initial cavity design. We also thank the members of the RADES collaboration for their valuable insights into the RF readout and discussions. 
We thank Eduard Chyhyrynets and the Istituto Nazionale di Fisica Nucleare di Legnaro for electropolishing the test cavity. We thank Aleksandr Zubtsovskii and the university of Siegen for coating the test cavity with NbN. We thanks Yuzhe Zhang for his help with the cryogenic infrastructure. 
This work would have not been possible without the ERC-Grant “LightAtTheLHC” as well as the continuous support from the PRISMA+ Cluster of Excellence at the University of Mainz.

\bibliography{MyBib}

\end{document}